\documentclass[conference]{IEEEtran}

\ifCLASSINFOpdf
\else
\usepackage[dvips]{graphicx}
\fi

\usepackage{amsmath,amssymb,amsfonts}
\usepackage{array}
\usepackage[caption=false,font=normalsize,labelfont=sf,textfont=sf]{subfig}
\usepackage{textcomp}
\usepackage{stfloats}
\usepackage{url}
\usepackage{verbatim}
\usepackage{graphicx}
\usepackage{cite}
\usepackage{xcolor}
\usepackage{orcidlink}
\usepackage[acronym,shortcuts]{glossaries}
\usepackage{bm}
\usepackage{relsize}
\usepackage{soul}
\usepackage{algorithm, algpseudocode}
\usepackage{algcompatible}
\usepackage{mathtools}
\usepackage{bm}
\usepackage{lipsum}
\usepackage{mathtools}
\usepackage{lipsum}
\usepackage[bottom]{footmisc}
\usepackage{tabularx}
\hypersetup{draft}

\def\BibTeX{{\rm B\kern-.05em{\sc i\kern-.025em b}\kern-.08em
T\kern-.1667em\lower.7ex\hbox{E}\kern-.125emX}}

\newcommand{\trans}[0]{^{\mathsf{T}}}

\newcommand{\herm}[0]{^{\mathsf{H}}}

\newcommand{\Real}[1]{\Re\{{#1}\}}
\newcommand{\Imag}[1]{\Im\{{#1}\}}

\newacronym{PCA}{PCA}{principal component analysis}
\newacronym{MDSM}{MDSM}{multi-domain sparse modulation}
\newacronym{P2P}{P2P}{point-to-point}
\newacronym{OTAC}{AirComp}{over-the-air computing}
\newacronym{TX}{TX}{transmitter}
\newacronym{RX}{RX}{receiver}
\newacronym{IoT}{IoT}{Internet-of-things}
\newacronym{AI/ML}{AI/ML}{artifitial intelligence/machine learning}
\newacronym{SDR}{SDR}{semi-definite relaxation}
\newacronym{EVD}{EVD}{eigenvalue decomposition}
\newacronym{GR}{GR}{Gaussian randomization}
\newacronym{SCA}{SCA}{successive convex approximation}
\newacronym{BnB}{BnB}{branch and bound}
\newacronym{QT}{QT}{quadratic transform}
\newacronym{RQ}{RQ}{Rayleigh quotient}
\newacronym{SOCP}{SOCP}{second-order cone programming}
\newacronym{CDF}{CDF}{cumulative distribution function}
\newacronym{UF}{UF}{uniform-forcing}
\newacronym{AP}{AP}{access point}
\newacronym{RSDR}{R-SDR}{regularized semi-definite relaxation}
\newacronym{R-SDR}{R-SDR}{regularized SDR}
\newacronym{flops}{flops}{floating point operations}
\newacronym{ED}{ED}{edge device}
\newacronym{SINR}{SINR}{signal to interference-plus-noise ratio}
\newacronym{SIC}{SIC}{successive interference cancellation}
\newacronym{CSI}{CSI}{channel state information}
\newacronym{LoS}{LoS}{line-of-sight}
\newacronym{NLoS}{NLoS}{non-LoS}
\newacronym{RPE}{RPE}{radar parameter estimation}
\newacronym{OTFS}{OTFS}{orthogonal time frequency space}
\newacronym{AFDM}{AFDM}{affine frequency division multiplexing}
\newacronym{CRLB}{CRLB}{Cram{\`e}r-Rao lower bound}
\newacronym{BCRLB}{BCRLB}{Bayesian Cram{\`e}r-Rao lower bound}
\newacronym{BBI}{BBI}{Bayesian bilinear inference}
\newacronym{AoA}{AoA}{angle-of-arrival}
\newacronym{SNR}{SNR}{signal-to-noise ratio}
\newacronym{ML}{ML}{maximum likelihood}
\newacronym{MIMO}{MIMO}{multiple-input multiple-output}
\newacronym{SIMO}{SIMO}{single-input multiple-output}
\newacronym{SISO}{SISO}{single-input single-output}
\newacronym{MUSIC}{MUSIC}{multiple signal classification}
\newacronym{MU}{MU}{multi-user}
\newacronym{ROOT-MUSIC}{ROOT-MUSIC}{ROOT multiple signal classification}
\newacronym{JCAS}{JCAS}{joint communication and sensing}
\newacronym{JCR}{JCR}{joint communications and radar}
\newacronym{ISAC}{ISAC}{integrated sensing and communications}
\newacronym{3D}{3D}{three-dimensional}
\newacronym{2D}{2D}{two-dimensional}
\newacronym{1D}{1D}{one-dimensional}
\newacronym{BF}{BF}{beamforming}
\newacronym{ROI}{ROI}{region of interest}
\newacronym{mmWave}{mmWave}{millimeter-wave}
\newacronym{MF}{MF}{matched-filter}
\newacronym{DD}{DD}{delay-Doppler}
\newacronym{SotA}{SotA}{state-of-the-art}
\newacronym{ULA}{ULA}{uniform linear array}
\newacronym{QAM}{QAM}{quadrature amplitude modulation}
\newacronym{ISFFT}{ISFFT}{inverse symplectic finite Fourier transform}
\newacronym{SFFT}{SFFT}{symplectic finite Fourier transform}
\newacronym{ISI}{ISI}{inter-symbol interference}
\newacronym{AWGN}{AWGN}{additive white Gaussian noise}
\newacronym{MSE}{MSE}{mean-squared-error}
\newacronym{LMMSE}{LMMSE}{linear minimum mean square error}
\newacronym{RMSE}{RMSE}{root mean square error}
\newacronym{ESPRIT}{ESPRIT}{estimation of signal parameters via rotational invariant techniques}
\newacronym{OFDM}{OFDM}{orthogonal frequency division multiplexing}
\newacronym{OCDM}{OCDM}{orthogonal chirp division multiplexing}
\newacronym{BS}{BS}{base station}
\newacronym{UE}{UE}{user equipment}
\newacronym{JCEDD}{JCEDD}{joint channel estimation and data detection}
\newacronym{PDA}{PDA}{probabilistic data association}
\newacronym{PMF}{PMF}{probability mass function}
\newacronym{PBiGaBP}{PBiGaBP}{parametric bilinear Gaussian belief propagation}
\newacronym{PBiGAMP}{PBiGAMP}{parametric bilinear generalized approximate message passing}
\newacronym{GaBP}{GaBP}{Gaussian belief propagation}
\newacronym{FT}{FT}{frequency-time}
\newacronym{DFT}{DFT}{discrete Fourier transform}
\newacronym{IDFT}{IDFT}{inverse discrete Fourier transform}
\newacronym{TD}{TD}{time domain}
\newacronym{wlg}{w.l.g.}{without loss of generality}
\newacronym{CP}{CP}{cyclic prefix}
\newacronym{DAF}{DAF}{discrete affine Fourier}
\newacronym{DAFT}{DAFT}{discrete affine Fourier transform}
\newacronym{IDAFT}{IDAFT}{inverse discrete affine Fourier transform}
\newacronym{CPP}{CPP}{\textit{chirp-periodic} prefix}
\newacronym{IDZT}{IDZT}{inverse discrete Zak transform}
\newacronym{DZT}{DZT}{discrete Zak transform}
\newacronym{P/S}{P/S}{parallel-to-serial}
\newacronym{S/P}{S/P}{serial-to-parallel}
\newacronym{SBL}{SBL}{sparse Bayesian learning}
\newacronym{MPA}{MPA}{message passing algorithms}
\newacronym{EM}{EM}{expectation maximization}
\newacronym{sIC}{soft IC}{soft interference cancellation}
\newacronym{soft RG}{soft RG}{soft replica generation}
\newacronym{BG}{BG}{belief generation}
\newacronym{SGA}{SGA}{scalar Gaussian approximation}
\newacronym{CLT}{CLT}{central limit theorem}
\newacronym{PDF}{PDF}{probability density function}
\newacronym{QPSK}{QPSK}{quadrature phase-shift keying}
\newacronym{ICI}{ICI}{inter-carrier interference}
\newacronym{BER}{BER}{bit error rate}
\newacronym{DoF}{DoF}{degrees-of-freedom}
\newacronym{VGA}{VGA}{vector Gaussian approximation}
\newacronym{FD}{FD}{full-duplex}
\newacronym{NMSE}{NMSE}{normalized mean square error}
\newacronym{KL}{KL}{Kullback-Leibler}
\newacronym{LASSO}{LASSO}{least absolute shrinkage and selection operator}
\newacronym{FP}{FP}{fractional programming}
\newacronym{CC}{CC}{communication-centric}
\newacronym{RC}{RC}{raised-cosine}
\newacronym{RRC}{RRC}{root raised-cosine}
\newacronym{6G}{6G}{sixth-generation}
\newacronym{V2X}{V2X}{vehicle-to-everything}
\newacronym{LEO}{LEO}{low-earth orbit}
\newacronym{I/O}{I/O}{input-output}
\newacronym{CE}{CE}{channel estimation}
\newacronym{ICC}{ICC}{integrated communication and computing}
\newacronym{ISCC}{ISCC}{integrated sensing, communications and computing}
\newacronym{PAM}{PAM}{pulse amplitude modulation}
\newacronym{iid}{i.i.d.}{independent and identically distributed}

\begin{document}

\title{From Theory to Reality: A Design Framework for\\
Integrated~\!Communication~\!and~\!Computing~\!Receivers\\[-1ex]} 

\author{\IEEEauthorblockN{Kuranage Roche Rayan Ranasinghe\textsuperscript{\orcidlink{0000-0002-6834-8877}}, Kengo Ando\textsuperscript{\orcidlink{0000-0003-0905-2109}} and Giuseppe Thadeu Freitas de Abreu\textsuperscript{\orcidlink{0000-0002-5018-8174}}}
\IEEEauthorblockA{\textit{School of Computer Science and Engineering, Constructor University, 28759 Bremen, Germany} \\
(kranasinghe,kando,gabreu)@constructor.university}\\[-6ex]
}

\maketitle

\begin{abstract}
We propose a novel flexible and scalable framework to design \ac{ICC} -- a.k.a. \ac{OTAC} -- receivers.
To elaborate, while related literature so far has generally focused either on theoretical aspects of \ac{ICC} or on the design of \ac{BF} algorithms for \ac{OTAC}, we propose a framework to design receivers capable of simultaneously detecting communication symbols and extracting the output of the \ac{OTAC} operation, in a manner that can: a) be systematically generalized to any nomographic function, b) scaled to a massive number of \acp{UE} and \acp{ED}, and c) support the multiple computation streams.
For the sake of illustration, we demonstrate the proposed method under a setting consisting of the uplink from multiple single-antenna \acp{UE}/\acp{ED} simultaneously transmitting communication and computing signals to a single multiple-antenna \ac{BS}/\ac{AP}.
The receiver, which seeks to detect all communication symbols and minimize the distortion over the computing signals, requires that only a fraction of the transmit power be allocated to the latter, therefore coming close to the ideal (but unattainable) condition that computing is achieved ``for free'', without taking resources from the communication system.
The design leverages the \ac{GaBP} framework relying only on element-wise scalar operations, which allows for its use in massive settings, as demonstrated by simulation results incorporating up to 200 antennas and 200 \acp{UE}/\acp{ED}. 
They also demonstrate the efficacy of the proposed method under all various loading conditions, with the performance of the scheme approaching fundamental limiting bounds in the under/fully loaded cases.
\end{abstract}


\glsresetall

\IEEEpeerreviewmaketitle

\vspace{-2ex}
\section{Introduction}
\label{sec:introduction}

\IEEEPARstart{T}{he} integration of functionalities such as sensing \cite{LiuJSAC_ISAC2022} and computing \cite{QiTWC2021} into communications systems is expected to be the key differential between the \ac{6G}  \cite{Cheng-XiangCST2023} and the current generation of wireless systems \cite{ShafiquACCESS20}.
However, while both sensing and computing are equally relevant in expanding the impact of wireless systems onto the modern way of living, it can be said that \ac{ISAC} has proven a somewhat easier task than \ac{OTAC}, as indicated by related literature which offers a fast-growing number of transceiver design approaches for \ac{ISAC} \cite{GaudioTWC2020, Ranasinghe_ICASSP_2024, Bemani_WCL_2024, RanasingheARXIV2024,Gonzalez_ProcIEEE_2024, RanasingheARXIV_blind_2024,Zhang_WC_2024, RexhepiARXIV2024}, while the focus of \ac{OTAC} contributions has remained so far on theoretical analysis \cite{NazerTIT07, LiuTWC20, Wang_arxiv_2024, QiaoLanWCL2020, QinWCL21} and \ac{BF} schemes \cite{ChenWCL18,FangSPAWC21, AndoCAMSAP2023}. 

In view of the above, this article proposes a novel \ac{GaBP}-based \cite{bickson2009gaussian, LiTSP2024, takahashi2018design} receiver design framework for \ac{ICC} in which both data symbols and computing signals are detected in order to yield effective joint communication and computing functionalities\footnotemark. 

To this end, we first formulate a system model in which the communication and computing signals are transmitted simultaneously by the \acp{UE} or \acp{ED}, both of which are to be detected by the receiving \ac{BS} or \ac{AP}.
We then derive the relevant message passing rules to extract the separate data symbol and computing streams via the \ac{GaBP} framework, enhanced with a closed-form combiner design to an optimization problem incorporating \ac{SIC}.

For the sake of illustration and simplicity of exposition, in this first article, the message passing rules are designed to extract the full set of individual elements of the computing signal, as opposed to traditional schemes which only estimate a specific nomographic function, as a step towards extracting the statistical prior on the aforementioned measurement data.
It can be understood from the derivations, however, that the message-passing rules for any desired nomographic function, such as those listed in \cite{perezneira2024waveformscomputingair}, can also be obtained provided that the corresponding prior distributions are derived, which is relegated to a follow-up work.

As a result of the strategy, the proposed framework can (in principle) be applied systematically to compute  any nomographic function and even to support the simultaneous computing of multiple computation streams.
In addition, as a consequence of the low complexity inherent to the \ac{GaBP} framework, the method can be scaled to massive setups, as demonstrated by simulation results shown for a system with up to 200 antennas at the \ac{BS}/\ac{AP} and 200 \acp{UE}/\acp{ED}.
As a bonus, simulation results also indicate that the approach is effective even if the power allocated to the computing signals is only a fraction of that allocated to communications symbols, bringing the overall method close to the ideal condition that computing is achieved ``for free'', without exploiting resources from the communication system.

\footnotetext{The integration of sensing into the \ac{ICC} framework considered in this manuscript, which would yield a more complete \ac{ISCC} problem, is rather trivial under the assumption that the communication waveforms can be used for sensing purposes via their reflected echoes, as demonstrated by related literature \cite{GaudioTWC2020, Ranasinghe_ICASSP_2024, Bemani_WCL_2024, RanasingheARXIV2024,Gonzalez_ProcIEEE_2024, RanasingheARXIV_blind_2024,Zhang_WC_2024, RexhepiARXIV2024}. 
However, due to space limitations, we leave the sensing component to be addressed directly in a follow-up work.}

The remainder of the article can be summarized as follows.
The system model is described in Section \ref{sec:system_model}.
The proposed method is then introduced in Section \ref{sec:proposed_method}.
Finally, the performance of both the communication and computing functionalities is evaluated in Section \ref{sec:performance}, in terms of \ac{BER} and \ac{MSE}, respectively.

\textit{Notation:} The following notation is used persistently throughout the manuscript.
Vectors and matrices are represented by lowercase and uppercase boldface letters, respectively;
$\mathbf{I}_M$ denotes an identity matrix of size $M$ and $\mathbf{1}_M$ denotes a column vector composed of $M$ ones; 
the vector norm and the absolute value of a scalar are respectively given by $\|\cdot\|$ and $|\cdot|$;
the transpose and hermitian operations follow the conventional form $(\cdot)\trans$ and $(\cdot)\herm$, respectively;
$\Re{\{\cdot\}}$, $\Im{\{\cdot\}}$ and  $\mathrm{min}(\cdot)$ represents the real part, imaginary part and the minimum operator, respectively.
Finally, $\sim \mathcal{N}(\mu,\sigma^2)$ and $\sim \mathcal{CN}(\mu,\sigma^2)$ respectively denotes the Gaussian and complex Gaussian distribution with mean $\mu$ and variance $\sigma^2$, where $\sim$ stands for ``is distributed as''.

\vspace{-0.5ex}
\section{System Model}
\label{sec:system_model}

\begin{figure}[t]
\centering
\includegraphics[width=0.9\columnwidth]{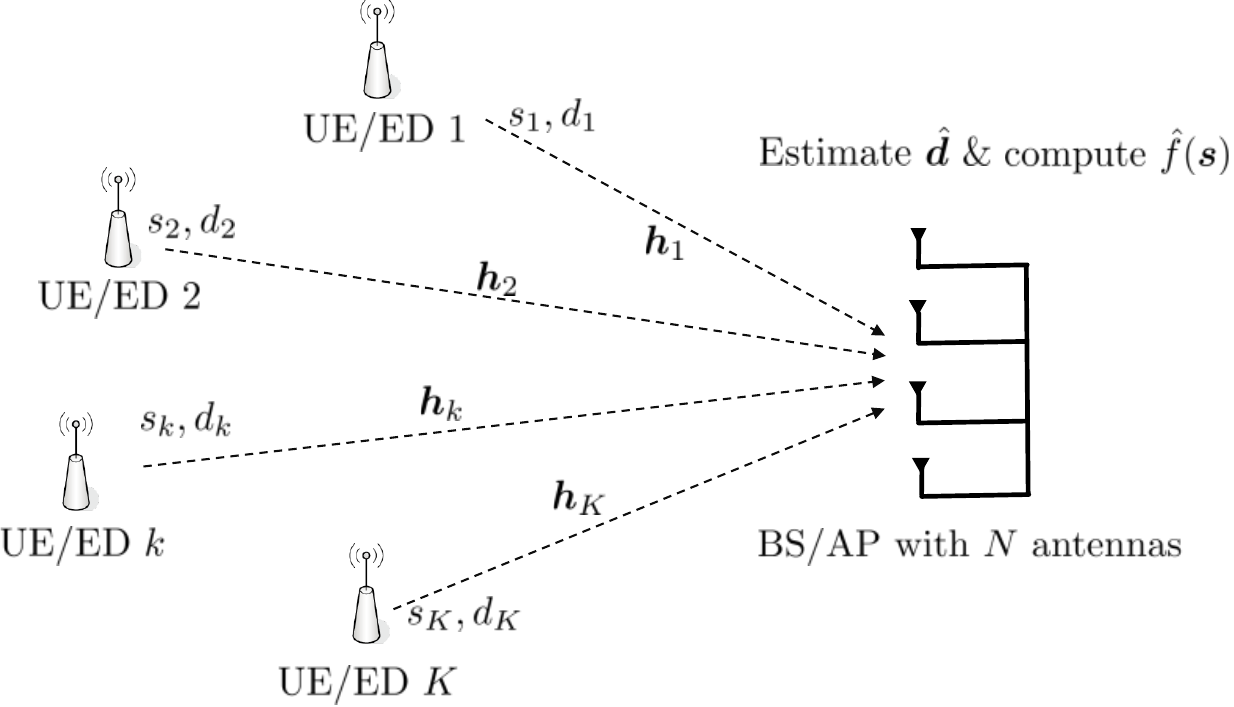}
\vspace{-1ex}
\caption{Illustration of a \ac{SIMO} \ac{ICC} system consisting of one \ac{BS}/\ac{AP} with $N$ antennas and $K$ single antenna \acp{UE}/\acp{ED}.}
\label{fig:system_model}
\vspace{-1ex}
\end{figure}

We consider an uplink multi-user \ac{SIMO} system composed of $K$ single-antenna \acp{UE}/\acp{ED} and one \ac{BS}/\ac{AP} equipped with $N$ antennas, as illustrated in Fig. \ref{fig:system_model}.
Under the assumption of perfect symbol synchronization amongst users, the received signal $\bm{y} \in \mathbb{C}^{N\times 1}$ at the \ac{BS}/\ac{AP} subjected to fading and noise is given by
\vspace{-0.5ex}
\begin{equation}
\label{eq:received_signal}
\bm{y} = \sum_{k=1}^K \bm{h}_k  {x}_k + \bm{w},
\vspace{-0.5ex}
\end{equation}
where the $\bm{h}_k \in \mathbb{C}^{N\times 1}$ denotes the channel vector between the $k$-th user and the \ac{BS}/\ac{AP} with each element $h_{n,k} \sim \mathcal{CN}(0,\sigma_h^2)$; ${x}_k\in \mathbb{C}$ is a concatenated transmit signal from $k$-th user and $\bm{w} \in \mathbb{C}^{N\times1}\sim \mathcal{CN}(0,\sigma^2_w\mathbf{I}_N)$ represents the \ac{AWGN}.

For the incorporation of both communication and computing functionalities into the system, the transmit signal can be decomposed and defined as a sum of its communication and computing parts, respectively, as
\vspace{-0.5ex}
\begin{equation}
\label{eq:transmit_sig_decomposition}
x_k \triangleq d_k +  \psi_k(s_k),
\vspace{-0.5ex}
\end{equation}
where $d_k \in \mathcal{D}$ and $s_k \in \mathbb{R}$\footnote{While the computing aspects usually incorporate the computation of a nomographic function composed of real variables \cite{LiuTWC20}, the extension to complex variables is trivial and holds for all the consequent derivations in the manuscript.} denote $k$-th user's modulated symbol for communication and computing, respectively, with $\mathcal{D}$ representing an arbitrary discrete constellation ($e.g.$ \ac{QAM}) with cardinality $D$; while $\psi_k(\cdot)$ denotes the pre-processing function for \ac{OTAC}.

Adhering to the computing aspect which requires the computation of an \ac{OTAC} target function at the \ac{BS}/\ac{AP}, we define $f(\bm{s})$ which can be described as
\begin{equation}
\label{eq:target_function_def}
f(\bm{s}) = \phi\left(\sum_{k=1}^K \psi_k(s_k)\right) \rightarrow \sum_{k=1}^K s_k,
\end{equation}
where $\phi$ represents the post-processing function for a general nomographic expression.
For the sake of simplicity, the arithmetic sum operation is chosen in this manuscript for the target function $f(\bm{s})$.

For later convenience, the received signal can now be reformulated in terms of matrix operations as
\begin{equation}
\bm{y} =  \bm{H}  \bm{x} + \bm{w} =\bm{H}  (\bm{d} + \bm{s}^\psi) + \bm{w},
\label{eq:received_signal_matrix}
\end{equation}
where the complex channel matrix $\bm{H} \triangleq [\bm{h}_1,\dots, \bm{h}_K] \in \mathbb{C}^{N\times K}$, the concatenated transmit signal $\bm{x} \triangleq [x_1,\dots, x_K]\trans \in \mathbb{C}^{K\times1}$, the data signal vector $\bm{d} \triangleq [d_1,\dots, d_K]\trans \in \mathbb{C}^{K\times1}$ and the computing signal vector $\bm{s}^\psi \triangleq [\psi_1(s_1),\dots, \psi_K(s_K)]\trans \in \mathbb{R}^{K\times1}$ are explicitly defined.

\section{Proposed Integrated Communications and Computing Receiver Design Framework}
\label{sec:proposed_method}
In this section, we consider a joint data and computing signal detection scheme termed \ac{ICC} using the well known \ac{GaBP} algorithm.
\Ac{wlg}, the modulation for communications is assumed to be \ac{QPSK}  and the $k$-th user's signal for computing is assumed to follow $s_k\sim\mathcal{N}(\mu_s,\sigma_s^2)$. 

%
Since the goal of \ac{OTAC} is to estimate a certain function that depends on the computing signal $\bm{s}$, the assumption that the mean of the elements $\mu_s$ is known is restrictive and has to updated iteratively with its corresponding estimates as detailed in the subsequent subsections.

\subsection{Joint Detection and Computing}

We now consider the joint estimation of the data symbols and the computing signals to fully recover the estimated data vector $\hat{\bm{d}} \in \mathbb{C}^{K \times 1}$ and the nomographic function $\hat{f}(\bm{s}) \in \mathbb{R}$ via the \ac{GaBP} technique.
As a consequence of the general element-wise structure of the \ac{GaBP}, we can formulate a bivariate set of message passing rules where the data signal vector $\bm{d}$ and the computing signal vector $\bm{s}$ can be estimated separately, each with their own Bayes-optimal denoisers for the best decoding/computing performance.

To that end, in order to derive the scalar \ac{GaBP} rules, let us first express equation \eqref{eq:received_signal_matrix} in an element-wise manner, dropping the superscript $\psi$ for convenience, as
\begin{equation}
y_n = \sum_{k=1}^K h_{n,k} \cdot d_k + \sum_{k=1}^K h_{n,k} \cdot s_k + w_n.
\label{eq:received_signal_elementwise}
\end{equation}

Next, consider the $i$-th iteration of the algorithm, and denote the soft replicas of the $k$-th communication and computing symbol with the $n$-th receive signal $y_n$ at the previous iteration respectively by $\hat{d}_{n,k}^{(i-1)}$ and $\hat{s}_{n,k}^{(i-1)}$. 

Then, the \acp{MSE} of these estimates computed for the $i$-th iteration are given by
\begin{subequations}
\begin{equation}
\hat{\sigma}^{2(i)}_{d:{n,k}} \triangleq \mathbb{E}_{d} \big[ | d - \hat{d}_{n,k}^{(i-1)} |^2 \big]= E_\mathrm{D} - |\hat{d}_{n,k}^{(i-1)}|^2, \forall (n,k),
\label{eq:MSE_d_k}
\end{equation}
\begin{equation}
\hat{\sigma}^{2(i)}_{s:{n,k}} \triangleq \mathbb{E}_{s} \big[ | s - \hat{s}_{n,k}^{(i-1)} |^2 \big], \forall (n,k),
\label{eq:MSE_s_k}
\end{equation}
\end{subequations}
where $\mathbb{E}_{d}$ and $\mathbb{E}_{s}$ respectively refers to expectation over all the possible symbols in the constellation $\mathcal{D}$ and the expectation over all the possible outcomes of $s \sim \mathcal{N}(\mu_s,\sigma^2_s)$ with $E_\mathrm{D}$ explicitly denoting the data constellation power.

\subsubsection{Soft Interference Cancellation} The objective of the \ac{sIC} stage at a given $i$-th iteration of the algorithm is to utilize the soft replicas $\hat{d}_{n,k}^{(i-1)}$ and $\hat{s}_{n,k}^{(i-1)}$ from a previous iteration to calculate the data- and computing-centric \ac{sIC} symbols $\tilde{y}_{d:n,k}^{(i)}$ and $\tilde{y}_{s:n,k}^{(i)}$ with their corresponding variances $\tilde{\sigma}_{d:n,k}^{2(i)}$ and $\tilde{\sigma}_{s:n,k}^{2(i)}$.

Therefore, exploiting equation \eqref{eq:received_signal_elementwise}, the \ac{sIC} symbols for both the data and computing signals are given as
\begin{subequations}
\begin{equation}
\label{eq:d_soft_IC}
\resizebox{0.48 \textwidth}{!}{$
\begin{aligned}
\tilde{y}_{d:n,k}^{(i)} &= y_{n} - \sum_{q \neq k} h_{n,q}\hat{d}_{n,q}^{(i)} - \sum_{k=1}^K h_{n,k}\hat{s}_{n,k}^{(i)}, \\[-2ex]
&= h_{n,k} d_k + \underbrace{\sum_{q \neq k} h_{n,q}(d_q - \hat{d}_{n,q}^{(i)}) + \sum_{k = 1}^{K} h_{n,k}(s_k - \hat{s}_{n,k}^{(i)})+ w_n}_\text{interference + noise term}, \\[-12ex]
\end{aligned}
$}
\end{equation}
\vspace{6ex}
\begin{equation}
\label{eq:s_soft_IC}
\resizebox{0.48 \textwidth}{!}{$
\begin{aligned}
\tilde{y}_{s:n,k}^{(i)} &= y_{n} - \sum_{q \neq k} h_{n,q}\hat{s}_{n,q}^{(i)} - \sum_{k=1}^K h_{n,k}\hat{d}_{n,k}^{(i)}, \\[-2ex]
&= h_{n,k} s_k + \underbrace{\sum_{q \neq k} h_{n,q}(s_q - \hat{s}_{n,q}^{(i)}) + \sum_{k = 1}^{K} h_{n,k}(d_k - \hat{d}_{n,k}^{(i)})+ w_n}_\text{interference + noise term}, \\[-12ex]
\end{aligned}
$}
\end{equation}
\vspace{6ex}
\end{subequations}

In turn, leveraging \ac{SGA} to approximate the interference and noise terms as Gaussian noise, the conditional \acp{PDF} of the \ac{sIC} symbols are given by
\begin{subequations}
\begin{equation}
\label{eq:cond_PDF_d}
p_{\tilde{\mathrm{y}}_{\mathrm{d}:n,k}^{(i)} \mid \mathrm{d}_{k}}(\tilde{y}_{d:n,k}^{(i)}|d_{k}) \propto \mathrm{exp}\left[ -\frac{|\tilde{y}_{d:n,k}^{(i)} - h_{n,k} d_{k}|^2}{\tilde{\sigma}_{d:n,k}^{2(i)}} \right],
\end{equation}
\begin{equation}
\label{eq:cond_PDF_s}
p_{\tilde{\mathrm{y}}_{\mathrm{s}:n,k}^{(i)} \mid \mathrm{s}_{k}}(\tilde{y}_{s:n,k}^{(i)}|s_{k}) \propto \mathrm{exp}\left[ -\frac{|\tilde{y}_{s:n,k}^{(i)} - h_{n,k} s_{k}|^2}{\tilde{\sigma}_{s:n,k}^{2(i)}} \right],
\end{equation}
\end{subequations}
with their conditional variances expressed as
\begin{subequations}
\begin{equation}
\label{eq:soft_IC_var_d}
\tilde{\sigma}_{d:n,k}^{2(i)} = \sum_{q \neq k} \left|h_{n,q}\right|^2 \hat{\sigma}^{2(i)}_{d:{n,q}} + \sum_{k = 1}^{K} \left|h_{n,k}\right|^2 \hat{\sigma}^{2(i)}_{s:{n,k}} + \sigma^2_w,
\end{equation}
\begin{equation}
\label{eq:soft_IC_var_s}
\tilde{\sigma}_{s:n,k}^{2(i)} = \sum_{q \neq k} \left|h_{n,q}\right|^2 \hat{\sigma}^{2(i)}_{s:{n,q}} + \sum_{k = 1}^{K} \left|h_{n,k}\right|^2 \hat{\sigma}^{2(i)}_{d:{n,k}} + \sigma^2_w.
\end{equation}
\end{subequations}

\subsubsection{Belief Generation} With the goal of generating the beliefs for all the data and computing symbols, we first exploit \ac{SGA} under the assumption that $N$ is a sufficiently large number and that the individual estimation errors in $\hat{d}_{n,k}^{(i-1)}$ and $\hat{s}_{n,k}^{(i-1)}$ are independent.

Therefore, as a consequence of \ac{SGA} and in hand of the conditional \acp{PDF}, the extrinsic \acp{PDF} are obtained as
\begin{subequations}
\begin{equation}
    \label{eq:extrinsic_PDF_d}
    \prod_{q \neq n} p_{\tilde{\mathrm{y}}_{\mathrm{d}:q,k}^{(i)} \mid \mathrm{d}_{k}}(\tilde{y}_{d:q,k}^{(i)}|d_{k}) \propto \mathrm{exp}\left[ - \frac{(d_k - \bar{d}_{n,k}^{(i)})^2}{\bar{\sigma}_{d:n,k}^{2(i)}} \right],
\end{equation}
\begin{equation}
    \label{eq:extrinsic_PDF_s}
    \prod_{q \neq n} p_{\tilde{\mathrm{y}}_{\mathrm{s}:q,k}^{(i)} \mid \mathrm{s}_{k}}(\tilde{y}_{s:q,k}^{(i)}|s_{k}) \propto \mathrm{exp}\left[ - \frac{(s_k - \bar{s}_{n,k}^{(i)})^2}{\bar{\sigma}_{s:n,k}^{2(i)}} \right],
\end{equation}
\end{subequations}
where the corresponding extrinsic means are defined as
\begin{subequations}
\begin{equation}
\label{eq:extrinsic_mean_d}
\bar{d}_{n,k}^{(i)} = \bar{\sigma}_{d:n,k}^{(i)} \sum_{q \neq n} \frac{h^*_{q,k} \cdot \tilde{y}_{d:q,k}^{(i)}}{ \tilde{\sigma}_{d:q,k}^{2(i)}},
\end{equation}
\begin{equation}
\label{eq:extrinsic_mean_s}
\bar{s}_{n,k}^{(i)} = \bar{\sigma}_{s:n,k}^{(i)}  \sum_{q \neq n} \frac{h^*_{q,k} \cdot \tilde{y}_{s:q,k}^{(i)}}{ \tilde{\sigma}_{s:q,k}^{2(i)}},
\end{equation}
\end{subequations}
with the extrinsic variances given by
\begin{subequations}
\begin{equation}
\label{eq:extrinsic_var_d}
\bar{\sigma}_{d:n,k}^{2(i)} = \left( \sum_{q \neq n} \frac{|h_{q,k}|^2}{\tilde{\sigma}_{d:q,k}^{2(i)}} \right)^{\!\!\!-1},
\end{equation}
\begin{equation}
\label{eq:extrinsic_var_s}
\bar{\sigma}_{s:n,k}^{2(i)} = \left( \sum_{q \neq n} \frac{|h_{q,k}|^2}{\tilde{\sigma}_{s:q,k}^{2(i)}} \right)^{\!\!\!-1}.
\end{equation}
\end{subequations}

\subsubsection{Soft Replica Generation} This stage involves the exploitation of the previously computed beliefs and denoising them via a Bayes-optimal denoiser to get the final estimates for the intended variables.
A damping procedure can also be incorporated here to prevent convergence to local minima due to incorrect hard-decision replicas.

Since the data symbols originate from a discrete \ac{QPSK} alphabet, \ac{wlg}, the Bayes-optimal denoiser is given as
\begin{equation}
\vspace{-1ex}
\hat{d}_{n,k}^{(i)}\! =\! c_d\! \cdot\! \bigg(\! \text{tanh}\!\bigg[ 2c_d \frac{\Real{\bar{d}_{n,k}^{(i)}}}{\bar{\sigma}_{d:{n,k}}^{2(i)}} \bigg]\!\! +\! j\text{tanh}\!\bigg[ 2c_d \frac{\Imag{\bar{d}_{n,k}^{(i)}}}{\bar{\sigma}_{{d}:{n,k}}^{2(i)}} \bigg]\!\bigg),\!\!
\label{eq:QPSK_denoiser}
\end{equation}
where $c_d \triangleq \sqrt{E_\mathrm{D}/2}$ denotes the magnitude of the real and imaginary parts of the explicitly chosen \ac{QPSK} symbols, with its corresponding variance updated as in equation \eqref{eq:MSE_d_k}.

Similarly, since the computing signal follows a Gaussian distribution, the denoiser with a Gaussian prior and its corresponding variance is given as \footnote{It is worth noting that if the post-processed computing signals have zero mean, the Gaussian denoiser will only depend on the extrinsic means and the computing symbol variance.}
\begin{subequations}
\begin{equation}
\label{eq:s_denoiser_mean}
\hat{s}_{n,k}^{(i)} = \frac{\sigma^2_s \cdot \bar{s}_{n,k}^{(i)} + \bar{\sigma}_{s:n,k}^{2(i)} \cdot \mu_s}{\bar{\sigma}_{s:n,k}^{2(i)} + \sigma^2_s},
\end{equation}
\begin{equation}
\label{eq:s_denoiser_var}
\hat{\sigma}_{s:n,k}^{2(i)} = \frac{\sigma^2_s \cdot \bar{\sigma}_{s:n,k}^{2(i)}}{\bar{\sigma}_{s:n,k}^{2(i)} + \sigma^2_s}.
\end{equation}
\end{subequations}

After obtaining $\hat{d}_{n,k}^{(i)}$ and $\hat{s}_{n,k}^{(i)}$ as per equations \eqref{eq:QPSK_denoiser} and \eqref{eq:s_denoiser_mean}, the final outputs are computed by damping the results with damping factors $0 < \beta_d,\beta_s < 1$ in order to improve convergence \cite{Su_TSP_2015}, yielding
\vspace{-1ex}
\begin{subequations}
\begin{equation}
\label{eq:d_damped}
\hat{d}_{n,k}^{(i)} = \beta_d \hat{d}_{n,k}^{(i)} + (1 - \beta_d) \hat{d}_{n,k}^{(i-1)},
\end{equation}
\begin{equation}
\label{eq:s_damped}
\hat{s}_{n,k}^{(i)} = \beta_s \hat{s}_{n,k}^{(i)} + (1 - \beta_s) \hat{s}_{n,k}^{(i-1)}.
\end{equation}
\end{subequations}

In turn, the corresponding variances $\hat{\sigma}^{2(i)}_{d:{n,k}}$ and $\hat{\sigma}^{2(i)}_{s:{n,k}}$ are first correspondingly updated via equations \eqref{eq:MSE_d_k} and \eqref{eq:s_denoiser_var}, respectively, and then damped via
\vspace{-1ex}
\begin{subequations}
\begin{equation}
\label{eq:MSE_d_m_damped}
\hat{\sigma}^{2(i)}_{d:{n,k}} = \beta_d \hat{\sigma}_{d:{n,k}}^{2(i)} + (1-\beta_d) \hat{\sigma}_{d:{n,k}}^{2(i-1)},
\end{equation}
\begin{equation}
\label{eq:MSE_s_m_damped}
\hat{\sigma}^{2(i)}_{s:{n,k}} = \beta_s \hat{\sigma}_{s:{n,k}}^{2(i)} + (1-\beta_s) \hat{\sigma}_{s:{n,k}}^{2(i-1)}.
\end{equation}
\end{subequations}

Finally, as a result of the conflicting dimensions, the consensus update can be obtained as
\vspace{-1ex}
\begin{subequations}
\begin{equation}
\label{eq:d_hat_final_est}
\hat{d}_{k} = \left( \sum_{n=1}^N \frac{|h_{n,k}|^2}{\tilde{\sigma}_{d:n,k}^{2(i_\text{max})}} \right)^{\!\!\!-1} \! \! \left( \sum_{n=1}^N \frac{h^*_{n,k} \cdot \tilde{y}_{d:n,k}^{(i_\text{max})}}{ \tilde{\sigma}_{d:n,k}^{2(i_\text{max})}} \right),
\end{equation}
\begin{equation}
\label{eq:s_hat_final_est}
\hat{s}_{k}^{(i)} = \Re \Bigg\{ \left( \sum_{n=1}^N \frac{|h_{n,k}|^2}{\tilde{\sigma}_{s:n,k}^{2(i)}} \right)^{\!\!\!-1} \! \! \left( \sum_{n=1}^N \frac{h^*_{n,k} \cdot \tilde{y}_{s:n,k}^{(i)}}{ \tilde{\sigma}_{s:n,k}^{2(i)}} \right) \Bigg\},
\end{equation}
\end{subequations}
where we take advantage of the fact that $\bm{s}$ is real.

\subsection{Closed-form \ac{OTAC} Combiner Design}

For the computation of a target function $\sum s_k$ at the \ac{BS}/\ac{AP}, let us first reformulate the combining of the residual signal leveraging equation \eqref{eq:received_signal_matrix} after \ac{SIC} of the estimated communication signal $\hat{\bm{d}}$ as
\vspace{-1ex}
\begin{equation}
  \vspace{-1ex}
\label{eq:Aircomp_SIC}
\hat{f}(\bm{s}) = \bm{u}\herm(\bm{y} - \bm{H}\hat{\bm{d}}) = \bm{u}\herm(\bm{H} (\bm{s} - \check{\bm{d}}) +\bm{w}),
\end{equation}
where $\bm{u} \in \mathbb{C}^{N\times1}$ denotes the combining vector, and we intrinsically define a data signal error vector\footnote{$\check{\bm{d}}$ is equivalent to the \ac{MSE} of the data signal vector $\hat{\sigma}^{2(i_\text{max})}_{d:{n,k}}$ averaged over all $n$ to satisfy the conflicting dimensions.} $\check{\bm{d}} \triangleq \hat{\bm{d}} - \bm{d} \in \mathbb{C}^{K \times 1}$.

Leveraging the above formulation, let us consider the optimization problem given by
\begin{equation}
\label{eq:min_problem}
\underset{\bm{u}\in\mathbb{C}^{N\times1}}{\mathrm{minimize}} \hspace{3ex} \| f(\bm{s}) - \hat{f}(\bm{s})  \|_2^2,
\end{equation}
where the objective function is defined as
\begin{align}
\label{eq:obj_func_def}
\| f(\bm{s}) \!-\! \hat{f}(\bm{s})  \|_2^2 \!\triangleq\! \| \mathbf{1}_{K}\trans \cdot\bm{s} \!-\! \bm{u}\herm(\bm{H} (\bm{s} \!-\! \check{\bm{d}}) +\bm{w})  \|_2^2.
\end{align}

Then, the closed-form solution for the combining vector can be derived as
\begin{equation}
\label{eq:u_def_precoder}
\bm{u} = (\bm{H}(\sigma_s^2\mathbf{I}_{K} + \boldsymbol{\Omega} )\bm{H}\herm+\sigma^2_w \mathbf{I}_{N})^{-1}\cdot \bm{H} \cdot (\sigma_s^2\mathbf{I}_{K}) \cdot \mathbf{1}_{K},
\end{equation}
where $\boldsymbol{\Omega} \triangleq  \mathbb{E} [\check{\bm{d}} \check{\bm{d}}\herm]$ is the final instantaneous variance of the data signal computed via the \ac{GaBP}.

\subsection{Joint Integrated Communication and Computing Design}

We now combine the low complexity \ac{GaBP} with the closed-form \ac{OTAC} combiner to estimate both the data signal $\hat{\bm{d}}$ and obtain the computing function $\hat{f}(\bm{s})$.
The complete pseudocode for the procedure is summarized in Algorithm \ref{alg:proposed_ICC}.

\begin{algorithm}[t]
\caption{Joint Data Detection \& \ac{OTAC} for Integrated  Communication and Computing Systems}
\label{alg:proposed_ICC}
\setlength{\baselineskip}{11pt}
\textbf{Input:} receive signal vector $\bm{y}\in\mathbb{C}^{N\times 1}$, complex channel matrix $\bm{H}\in\mathbb{C}^{N\times K}$, maximum number of iterations $i_{\max}$, data constellation power $E_\mathrm{D}$, noise variance $\sigma^2_w$, computing signal variance $\sigma^2_s$ and damping factors $\beta_d,\beta_s$. \\
\textbf{Output:} $\hat{\bm{d}}$ and $\hat{f}(\bm{s})$ 
\vspace{-2ex} 
\begin{algorithmic}[1]  
\STATEx \hspace{-3.5ex}\hrulefill
\STATEx \hspace{-3.5ex}\textbf{Initialization}
\STATEx \hspace{-3.5ex} - Set iteration counter to $i=0$ and amplitudes $c_d = \sqrt{E_\mathrm{D}/2}$.
\STATEx \hspace{-3.5ex} - Set initial data estimates to $\hat{d}_{n,k}^{(0)} = 0$ and corresponding 
\STATEx \hspace{-2ex} variances to $\hat{\sigma}^{2(0)}_{d:{n,k}} = E_\mathrm{D}, \forall n,k$.
\STATEx \hspace{-3.5ex} - Set initial computing signal estimates to $\hat{s}_{n,k}^{(0)} = 0$ and 
\STATEx \hspace{-2ex} corresponding variances to $\hat{\sigma}^{2(0)}_{s:{n,k}} = \sigma_s^2, \forall n, k$.
\STATEx \hspace{-3.5ex} - Set $\mu_s^{(0)} = 0$.
\STATEx \hspace{-3.5ex}\hrulefill
\STATEx \hspace{-3.5ex}\textbf{for} $i=1$ to $i_\text{max}$ \textbf{do}
\STATEx \textbf{Communication and Computing Update}: $\forall n, k$
\STATE Compute \ac{sIC} data signal $\tilde{y}_{d:{n,k}}^{(i)}$ and its corresponding variance $\tilde{\sigma}^{2(i)}_{d:{n,k}}$ from equations \eqref{eq:d_soft_IC} and \eqref{eq:soft_IC_var_d}.
\STATE Compute \ac{sIC} computing signal $\tilde{y}_{s:{n,k}}^{(i)}$ and its corresponding variance $\tilde{\sigma}^{2(i)}_{s:{n,k}}$ from equations \eqref{eq:s_soft_IC} and \eqref{eq:soft_IC_var_s}.
\STATE Compute extrinsic data signal belief $\bar{d}_{n,k}^{(i)}$ and its corresponding variance $\bar{\sigma}_{d:{n,k}}^{2(i)}$ from equations \eqref{eq:extrinsic_mean_d} and \eqref{eq:extrinsic_var_d}.
\STATE Compute extrinsic computing signal belief $\bar{s}_{n,k}^{(i)}$ and its corresponding variance $\bar{\sigma}_{s:{n,k}}^{2(i)}$ from eqs. \eqref{eq:extrinsic_mean_s} and \eqref{eq:extrinsic_var_s}.
\STATE Compute denoised and damped data signal estimate $\hat{d}_{n,k}^{(i)}$ from equations \eqref{eq:QPSK_denoiser} and \eqref{eq:d_damped}.
\STATE Compute denoised and damped data signal variance $\hat{\sigma}_{d:{n,k}}^{2(i)}$ from equations \eqref{eq:MSE_d_k} and \eqref{eq:MSE_d_m_damped}.
\STATE Compute denoised and damped computing signal estimate $\hat{s}_{n,k}^{(i)}$ from equations \eqref{eq:s_denoiser_mean} and \eqref{eq:s_damped}.
\STATE Compute denoised and damped computing signal variance $\hat{\sigma}_{s:{n,k}}^{2(i)}$ from equations \eqref{eq:s_denoiser_var} and \eqref{eq:MSE_s_m_damped}.
\STATE Compute $\hat{s}_k^{(i)}, \forall k$ using equation \eqref{eq:s_hat_final_est}.
\STATE Update $\mu_s^{(i)} = \frac{1}{K} \sum_{k=1}^K \hat{s}_k^{(i)}$.

\STATEx \hspace{-3.5ex}\textbf{end for}
\STATEx \hspace{-3.5ex}\textbf{Communication and Computing Consensus}: 
\STATE Calculate $\hat{d}_k, \forall k$ (equivalently $\hat{\bm{d}}$) using equation \eqref{eq:d_hat_final_est}. 
\STATE Compute $\bm{u}$ from equation \eqref{eq:u_def_precoder}.
\STATE Compute $\hat{f}(\bm{s})$ from equation \eqref{eq:Aircomp_SIC}.

\end{algorithmic}
\end{algorithm}
\vspace{-1ex}

\section{Performance Assessment}
\label{sec:performance}

For all the numerical simulations carried out, the total transmit power is held constant at $1$ with $99\%$ of the power allocated to the data signal and $1\%$ of the power allocated to the computing signal to minimize the distortion to the overall transmit waveform, allowing for the exploitation of all the \acp{DoF} in waveform design for communications and/or sensing.
In addition, the computing signal is assumed to follow $s_k\sim\mathcal{N}(0,\sigma_s^2)$ and the channel coefficients follow $h_{n,k} \sim \mathcal{CN}(0,1)$.
Similarly, the algorithmic parameters are set as $\beta_d = 0.5$, $\beta_s = 0.8$ and $i_\text{max} = 30$.

\subsection{Numerical Results}

\begin{figure}[b]
    \centering
    \includegraphics[width=\columnwidth]{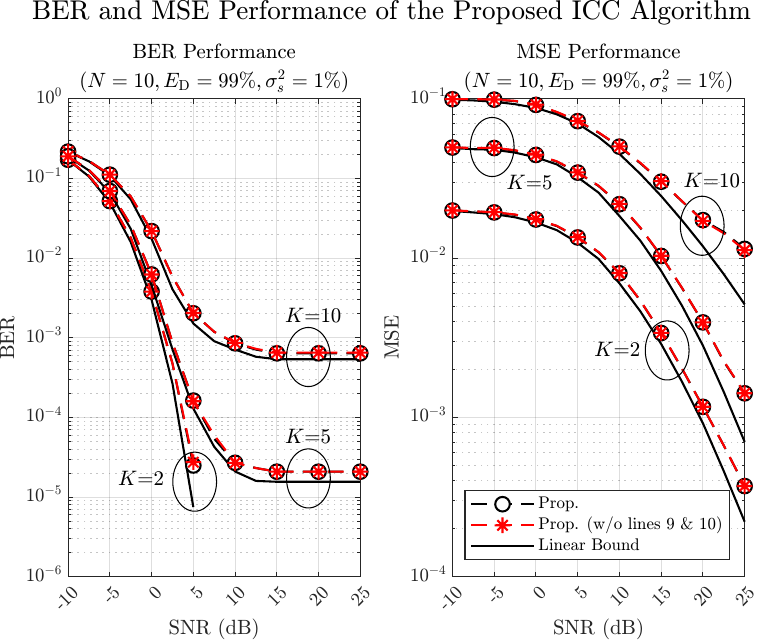}
    \caption{\ac{BER} and \ac{NMSE} performance of the proposed algorithm for the overloaded, underloaded and fully-loaded scenarios.}
    \label{fig:BER_MSE_plot_1}
    \vspace{-2ex}
\end{figure}

For the numerical evaluation of our proposed method, we first consider a typical uplink system composed of a \ac{BS}/\ac{AP} with $N = 10$ antennas servicing a varying number of single-antenna users that are categorized into the underloaded, fully loaded and overloaded scenarios with $K = 2$, $K = 5$ and $K = 10$ users, respectively\footnote{Notice that as a result of the bivariate estimation carried out, the fully loaded case is actually when $N=10$ and $K=5$ since a total of $2K$ variables in the form of data and computing symbols need to be estimated from $N$ factor nodes.}.   

Consequently, Fig. \ref{fig:BER_MSE_plot_1} now showcases \ac{BER} and \ac{MSE} results for the estimation of $\hat{\bm{d}}$ and $\hat{f}(\bm{s})$, respectively, for all three aforementioned cases as follows: the linear bound is obtained by running the proposed algorithm under the assumption that one of the estimates is perfectly known ($i.e.,$ for the case of data detection, the computing signals are assumed to be known and vice-versa); the result illustrated by the white circles is computed by running the proposed algorithm as it is and finally, the red dashed line presents the results under the assumption that the computing signals follow $s_k\sim\mathcal{N}(0,\sigma_s^2)$, leading to $\mu_s = 0$.

\begin{figure}[t]
    \centering
    \includegraphics[width=\columnwidth]{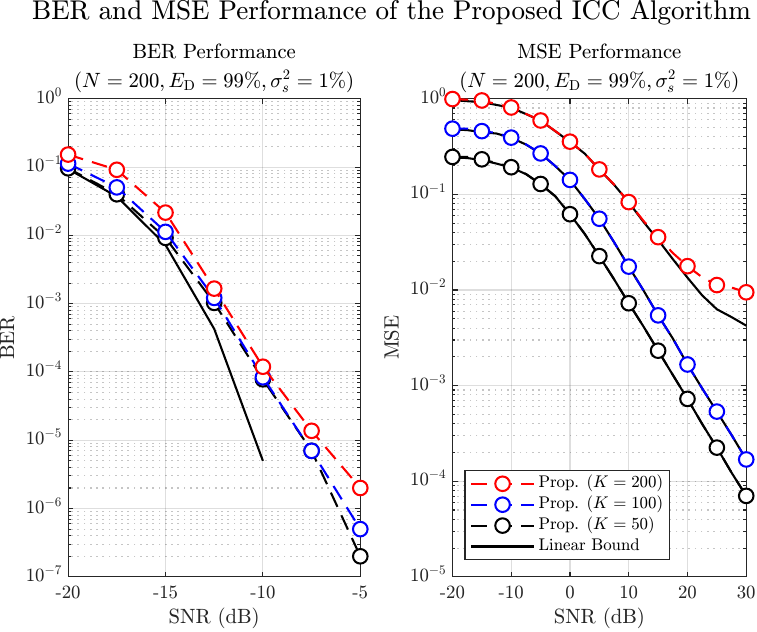}
    \caption{\ac{BER} and \ac{NMSE} Performance of the proposed algorithm in massive systems for the overloaded, underloaded and fully-loaded scenarios.}
    \label{fig:BER_MSE_plot_2}
    \vspace{-2ex}
\end{figure}

As demonstrated by the results, we see that all the variations, with more emphasis on the fully and underloaded cases of the proposed algorithm approaches the linear bound in terms of \ac{BER} with a slightly higher \ac{BER} seen by the high performance result.
In addition, although the fully and overloaded scenarios experience an error flow for the \acp{BER} at higher \acp{SNR}, these effects can be overcome with larger system sizes as seen from our next result in Fig. \ref{fig:BER_MSE_plot_2} and can also be mitigated via methods such as channel coding and iterative damping \cite{Su_TSP_2015} for lower system sizes.

Next, we consider a similar system setup but with a much larger number of $N = 200$ antennas servicing a varying number of single-antenna users that are also categorized into the underloaded, fully loaded and overloaded scenarios with $K = 50$, $K = 100$ and $K = 200$ users, respectively, to demonstrate the efficacy of the proposed algorithm for enabling \ac{ICC} in massive scenarios.

As seen in Fig. \ref{fig:BER_MSE_plot_2}, the \ac{BER} improves significantly with a larger system which can also be attributed to the strengthening of the \ac{SGA} utilized in the belief generation stage with the best performance seen by underloaded systems as expected.
In addition, considering the \ac{MSE} performance for computing, we see that the under and fully loaded scenarios reach the appropriate bounds in terms of performance with the overloaded case not too far behind as well, with the distinctions in the variations explained by the increasing variance with a larger $K$ as opposed to the \ac{BER} results.

\subsection{Complexity Analysis}

Due to the low complexity nature of the \ac{GaBP} framework, the resulting per-iteration computational complexity order for the initial communication and computing update is $\mathcal{O}(NK)$ which is linear on the number of element-wise computations performed.
However, since the final closed-form computing combiner on line 12 of Algorithm \ref{alg:proposed_ICC} requires a matrix inversion of size $N \times N$, the computational complexity order for the latter part is $\mathcal{O}(N^3)$.

\section{Future/Follow-up Work}

\subsection{Generalized Nomographic Functions}

The following are some distinct nomographic functions whose prior distributions can be exploited for \ac{ICC}, under the assumption that the individual elements are \ac{iid} and $s_k\sim\mathcal{N}(\mu_s,\sigma_s^2)$.

\begin{table}[t]
    \centering
    \caption{Examples of some nomographic functions and their priors.}
    \vspace{-1ex}
    \begin{tabular}[H]{|c|c|c|}
    \hline 
    \bf Description &\bf $f(\bm{s})$ &\bf Prior Distribution\\[0.5ex]
    \hline\hline 
    Arithmetic sum & $\sum_{k=1}^K s_k$ & $\sim \mathcal{N}(K\mu_s, K\sigma_s^2)$ \\[0.5ex]
    \hline
    Arithmetic product& $\prod_{k=1}^K s_k$ & \cite{gauss_prod} \\[0.5ex]
    \hline
    Max operation & $\underset{k}{\text{max}}\{ s_k \}$ & \cite{book_order_Statistics} \\[0.5ex]
    \hline
    Min operation & $\underset{k}{\text{min}}\{ s_k \}$ & \cite{book_order_Statistics} \\[0.5ex]
    \hline
    \end{tabular}
    \label{table:priors}
    \vspace{-2ex}
\end{table}

We also remark that if there exists some non-trivial nomographic function whose prior distributions cannot be computed analytically, they can, for example, be approximated via techniques using Gaussian mixtures or Jacobian methods \cite{AbreuTCom2008}.

\subsection{Multi-stream Computation}

Nomographic functions from the same/different streams of signals can also be computed together or separately via the proposed method, provided that there exists some prior distribution with the key advantage being the increase of only one incorporated variable to be estimated in the belief propagation framework.

\section{Conclusion}
\label{sec:conclusion}

In this manuscript, we proposed a novel framework for the design of practical \ac{ICC} receivers, with an emphasis on the flexibility and scalability of the systems.
Moreover, the resulting receiver framework, which seeks to detect all communication symbols and minimize the distortion over the computing signals, requires that only a fraction of the transmit power be allocated to the latter, therefore coming close to the ideal condition that computing is achieved without sacrificing any communications capability.
The key scalability enabler in this case is the utilized \ac{GaBP} structure relying only on element-wise scalar operations, enabling its use in massive settings, as demonstrated by simulation results incorporating up to 200 antennas and 200 \acp{UE}/\acp{ED}. 
Finally, further numerical simulations also demonstrate the efficacy of the proposed method under all various loading conditions, where the performance of the scheme approaches fundamental limiting bounds in under/fully loaded cases with a very low complexity.


\end{document}